\documentclass[reprint,
superscriptaddress,
 amsmath,amssymb,amsfonts,
 aps,
]{revtex4-1}
\usepackage[dvipsnames]{xcolor}
\usepackage{graphicx}
\usepackage{dcolumn}
\usepackage{bm}

\begin{document}


\title{Multipoint Bootstrap I: Light-Cone Snowflake OPE and the WL Origin.}

\author{Carlos Bercini}
\affiliation{ICTP South American Institute for Fundamental Research, IFT-UNESP,\\ São Paulo 01140-070, Brazil}

\author{Vasco Gon\c{c}alves}%
\affiliation{ICTP South American Institute for Fundamental Research, IFT-UNESP,\\ São Paulo 01140-070, Brazil}
\affiliation{Centro de Física do Porto e Departamento de Física e Astronomia,\\Faculdade de Ciências da Universidade do Porto, Porto 4169-007, Portugal}

\author{Pedro Vieira}
 \affiliation{Perimeter Institute for Theoretical Physics,\\ Waterloo, Ontario N2L 2Y5, Canada}
 \affiliation{ICTP South American Institute for Fundamental Research, IFT-UNESP,\\ São Paulo 01140-070, Brazil}

\begin{abstract}
We initiate an exploration of the conformal bootstrap for $n>4$ point correlation functions. Here we bootstrap correlation functions of the lightest scalar gauge invariant operators in planar non-abelian conformal gauge theories as their locations approach the cusps of a null polygon. For that we consider consistency of the OPE in the so-called snowflake channel with respect to cyclicity transformations which leave the null configuration invariant. For general non-abelian gauge theories this allows us to strongly constrain the OPE structure constants of up to three large spin $J_j$ operators (and large polarization quantum number $l_{j}$) to all loop orders. In $ \mathcal{N}=4$ we fix them completely through the duality to null polygonal Wilson loops and the recent origin limit of the hexagon explored by Basso, Dixon and Papathanasiou. 
\end{abstract}

\maketitle

\section{Introduction}
The \textit{numerical} conformal bootstrap \cite{numerics} is by now a well established physics tool, often leading to the best determination of critical exponent relevant for real world experiments. The so called \textit{analytical} conformal bootstrap explores a corner of the bootstrap -- usually the very Lorentzian domain -- to derive universal analytical results for general conformal field theories or general conformal gauge theories. 

A beautiful example is the work of Alday and Bissi~\cite{Alday:2013cwa}. They found the planar $\mathcal{N}=4$ structure constant between the lightest single trace scalars of the theory and the leading twist single trace operators with large spin $J$ as
\begin{equation}
\hat{C}^{\circ \circ \bullet}\simeq \mathcal{N}^{\circ \circ \bullet} \,  \Gamma\big(1-\tfrac{1}{2} \gamma\big) e^{-\frac{1}{2}(f\ln 2+g)\ln J}
\label{ABBeq}
\end{equation}
Here and below the hat in $\hat{C}$ stands for the all loop structure constant normalized by the tree level result and $\mathcal{N}^{\circ \circ \bullet}$ is a coupling dependent (but spin independent) normalization constant (which bootstrap arguments will always be insensitive to), \cite{footnoteCNotation}. The functions $f$ and $g$ are, respectively, the cusp and collinear anomalous dimensions. They show up in the 
 the anomalous dimension $\gamma$ of the leading twist operators at large spin $J$  which exhibits logarithmic scaling \cite{Korchemsky} and behave as 
\begin{equation}
\gamma \simeq f(\lambda) \ln(J) + g(\lambda)\,.
\label{anomalousDimension}
\end{equation}
in any gauge theory. Although extensively checked in $\mathcal{N}=4$ where we have abundant perturbative data, the derivation is really a bootstrap one and as such the prediction (\ref{ABBeq}) is actually expected to hold for any large $N$ conformal gauge theory.
 
In principle, the study of \textit{all} four point correlation functions contains information about all the CFT data. In practice most bootstrap studies focus on a single correlator or on a coupled system involving a few four point correlators. Here we propose to use \textit{higher} point functions as well. Since they contain, in their OPE, infinitely many four point correlation functions, we can expect that a great deal of information can be extracted from them.

We will obtain several new analytic results akin to (\ref{ABBeq}) but involving several spinning operators, namely (\ref{CJJScalar}) and (\ref{CJJJFinal}). Our results will be valid for general conformal gauge theories in the planar limit but our testing ground will once again be planar~$\mathcal{N}=4$ SYM.
Better understanding the various dualities relating scalar correlation functions, null polygonal Wilson loops (WL), gluon scattering amplitudes and large spin correlation functions \cite{d1,d2,d3,d4,Drummond:2007au,nullOPE} in this gauge theory is another motivation for these bootstrap exercises. Correlation functions correspond to AdS closed string scattering and are nicely built out of integrable hexagons~\cite{hexagons} while null WL can be depicted as open string partition functions and can be constructed out of integrable pentagons~\cite{Basso:2013vsa}. We hope these explorations will lead to a unifying description of both. 

\section{Kinematics} \label{Kin}
The OPE of two operators on the light cone was written around 50 year ago \cite{Ferrara:1974nf}:
\begin{align}
\mathcal{O}(x_1)\mathcal{O}(x_2) \sim& \frac{C(2x_{12} \cdot \partial_{\epsilon})^{J}}{(x_{12}^2)^{\frac{2\Delta_{\phi}-(\Delta-J)}{2}}} \label{eq:lightconeOPEE}\\
&\!_1F_1(\frac{\Delta+J}{2},\Delta+J,x_{21}\cdot\partial_{x_1})\mathcal{O}_{\Delta,J}(x_1,\epsilon) \nonumber \,.
\end{align}
where $C$ is the OPE coefficient. 
Here $\epsilon$ is a polarization vector and we should only keep leading terms in $x_{12}^2$. This operator identity can be used to derive integral representations for  conformal blocks by applying  it once, twice or three times in a $4$,$5$ and $6$ point correlator, respectively. 
\begin{figure}[t]
	\centering 
	\includegraphics[width=0.9\linewidth]{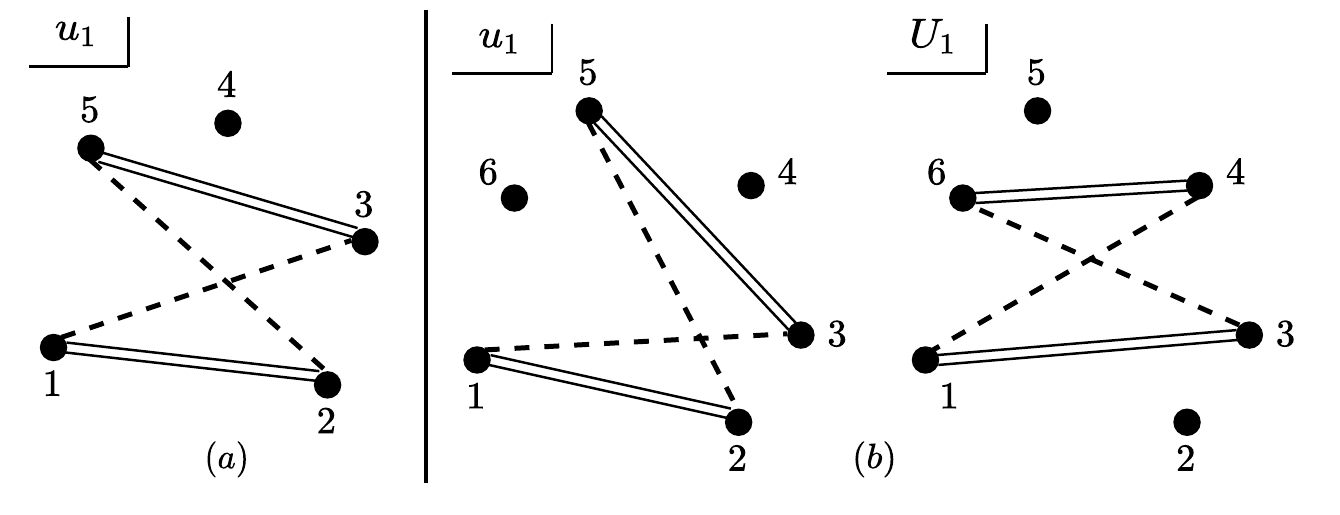}
	\caption{{\textbf a)} We pick the five independent cross-ratios for the 5pt function, as $u_1$ represented here plus its four cyclic images. {\textbf b)} For the 6pt function, we choose the nine independent cross-ratios as $u_1$ ($U_1$) represented here plus its five (two) cyclic images.}
	\label{5pt cross-ratios}
\end{figure}
After these multiple OPEs end up with the conformal block as the action of a few $\!_1F_1$ operators on a spinning three point function which is turn is completely fixed by conformal symmetry \cite{Costa:2011mg}:
\begin{align}
&\langle \mathcal{O}(x_1,\epsilon_1)\dots\mathcal{O}(x_3,\epsilon_3)\rangle = \sum_{l_{ij}}C_{J_1J_2J_3}^{l_{12}l_{13}l_{23}}\times\nonumber\\
&\frac{V_{1,23}^{J_1-l_{12}-l_{13}}V_{2,\textcolor{black}{31}}^{J_2-l_{12}-l_{23}}V_{3,12}^{J_3-l_{13}-l_{23}}H_{12}^{l_{12}}H_{\textcolor{black}{31}}^{l_{13}}H_{23}^{l_{23}}}{(x_{12}^2)^{\frac{\tau_1+\tau_2-\tau_3}{2}}(x_{13}^2)^{\frac{\tau_1+\tau_3-\tau_2}{2}}(x_{23}^2)^{\frac{\tau_2+\tau_3-\tau_1}{2}}} \label{eq:3ptStructureSpin}
\end{align}
where $\epsilon_i$ are the polarization vectors, \textcolor{black}{$\tau_i=\Delta_i+J_i$ are the conformal spin} of the operators and the $V$ and $H$ tensors are recalled in appendix \ref{BlocksAppendix}. The integers $l_{ij}$ parametrize the various possible tensor structures where the polarizations vectors show up and we simply call them \textit{polarizations}. Their range is such that all exponents in~(\ref{eq:3ptStructureSpin}) are non-negative. 
(The five-point block only depends on two spins $J_1,J_2$ and a single non-zero polarization ${ l}\equiv l_{12}$.) Finally, we use the standard integral representation for each $\!_1F_1$ to obtain a useful expression for the conformal blocks
\begin{equation}
F_n = \texttt{kin}_n \int_0^1 [dy_1]\dots[dy_{n-3}]\texttt{int}_n
\label{eq:Blockslightcone}
\end{equation}
where the explicit expressions for the kinematical prefactors, integrands and measure are summarized in appendix \ref{BlocksAppendix}.

\begin{figure}[t]
	\centering 
	\includegraphics[width=0.8\linewidth]{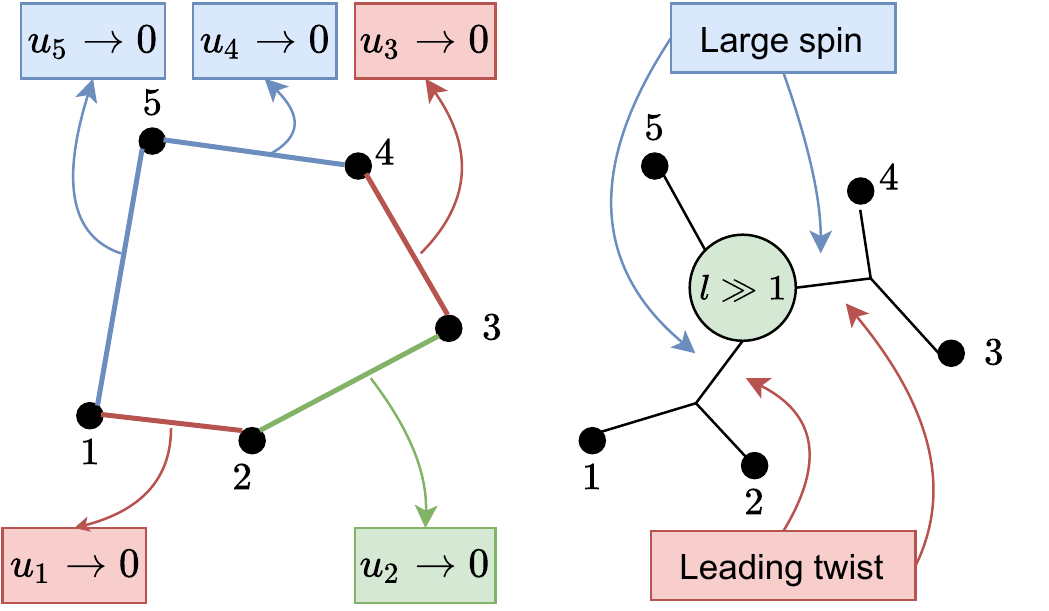}
	\caption{Each time we take consecutive points $x_i$ and $x_{i+1}$ to become null separated, we have $u_i \to 0$. We can thus construct a full null pentagon sequentially: $u_{1,3} \to 0 $ projects into leading twist;~$u_{4,5} \to 0 $ projects into large spin; finally~$u_{2} \to 0$ is controlled by structure constants with large $l$. }
	\label{5pt limits}
\end{figure}

Correlation functions of 5 and 6 points depend on 5 and 9 independent cross-ratios respectively. For the 5pt function, a convenient choice are the cross-ratios~\cite{Alday:2010zy}
\begin{equation}
u_1 \equiv \frac{x_{12}^2 x_{35}^2}{x_{13}^2 x_{25}^2} \,, \qquad x_{ij}^2 \equiv (x_i-x_j)^2 \label{u1BDS}
\end{equation}
while all other four cross-ratios are trivially obtained by simply shifting the indices here, 
\begin{equation}
u_{i} \equiv \left. u_{i-1}  \right|_{x_i \to x_{i+1}} \label{uiBDS}\,,
\end{equation}
where $i=2,3,4,5$, see figure \ref{5pt cross-ratios}. These cross-ratios contain a single nearest neighbor distance~($x_{12}$ in (\ref{u1BDS})). 
The light cone OPE in the channel 12 and 34 channel projects into leading twist contributions and can be achieved by taking $u_1, u_3\rightarrow 0$, see figure~\ref{5pt limits}. For  $x_{15}^2,x_{45}^2\rightarrow 0$ (or $u_4,u_5\to 0$) we are further dominated by large spin operators. Then the leading twist blocks simplify dramatically into a simple product of Bessel functions
\begin{equation}
F_5 \simeq \frac{(1-u_2)^l}{\pi} \prod_{i=1}^2\frac{2^{2J_i+\gamma_i}J_i^\frac{1}{2}}{u_{1-i}^{-\frac{2l+\gamma_{i+1}}{4}}}K_{l+\frac{\gamma_{i+1}}{2}}\left(2J_i\sqrt{u_{1-i}}\right)\label{Block5pt}
\end{equation}
At this point, the polarization $l$ is still finite; in the full null pentagon limit where all $x_{i,i+1}^2 \to 0$ (i.e. all $u_i \to 0$)  we project into the limit of large polarization further simplifying these Bessel functions into simple exponentials, see (\ref{F5check}).

\begin{figure}[t]
\centering 
	\includegraphics[width=0.8\linewidth]{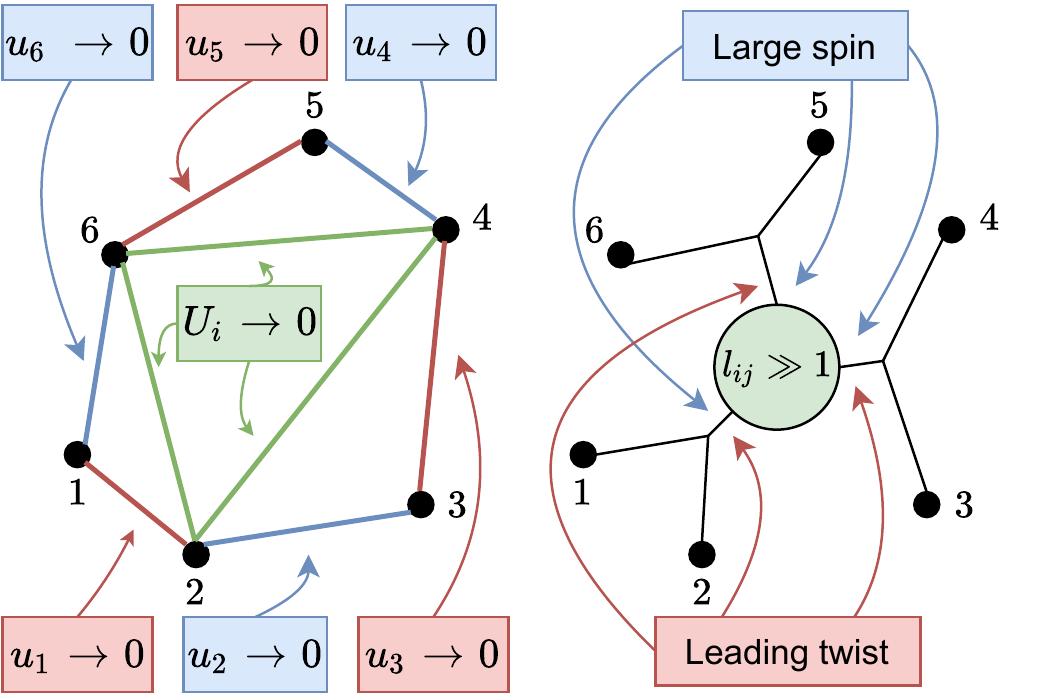}
	\caption{Consecutive 6pt limits: $u_{1,3,5} \to 0 $ projects into leading twist; $u_{2,4,6} \to 0 $ projects into large spin; finally $U_{i} \to 0$ is controlled by structure constants with large $l_{ij}$.}
	\label{6pt limits}
\end{figure}

For six point functions
we define 6 cross-ratios $u_j$ through (\ref{u1BDS}) and (\ref{uiBDS}) where $i$ now runs up to $6$ in the second relation. The remaining 3 cross-ratios $(U_1, U_2, U_3)$ which parametrize the six point functions are
\begin{equation}
U_1 \equiv \frac{x_{13}^2 x_{46}^2}{x_{14}^2 x_{36}^2} \, , \,\,\,\, U_{2} = \left. U_{1}  \right|_{x_i \to x_{i+1}}   \, , \,\,\,\, U_{3} = \left. U_{2}  \right|_{x_i \to x_{i+1}}  \,, \nonumber
\end{equation}
see figure \ref{5pt cross-ratios}. Again, we can explore a null hexagon limit by taking all consecutive points to become null separated. First we take $12$, $34$ and $56$ to become null by taking $u_1,u_3, u_5$ to zero; this projects into leading twist operators (with finite spins $J_1,J_2,J_3$) in these three OPE channels. Next we take $23$, $45$ and $61$ to become null by taking $u_2,u_4,u_6$ to zero; this projects into large spins $J_i \gg 1$, see figure~\ref{6pt limits}. We consider a corner of this limit where we also take $24$, $46$ and $62$ to become null, see figure ~\ref{6pt limits}. If they become null with the same rate the conformal blocks simplify considerably in this limit and are dominated by large spin $J_i$
\begin{align}
F_6\simeq &\prod_{i=1}^3 \frac{2^{2J_i+\gamma_i}J_i^{\frac{1}{2}}K_{l_{i-1}-l_{i}+\frac{\gamma_{i}-\gamma_{i-1}}{2}}(2J_{i+1}\sqrt{{U}_{2-i}})}{U_{2-i}^{-\frac{\gamma_{i}-\gamma_{i-1}+2l_{i}-2l_{i-1}}{4}}({U_{2-i}}-u_{2i})^{-l_{i-1}}}
\label{eq:Block6ptlightcone}
\end{align}
where $l_{1}=l_{23},l_{2}=l_{13},l_{3}=l_{12}$ are kept finite and where all indices are understood in cyclic sense with respect to their range, so that $U_{-2}=U_{3-2}=U_{1}$ and so on. We denote this limit -- where all cross-ratios vanish with the same rate -- as the \textit{generalized origin} limit.
Now, if we take the diagonals to approach the null limit faster than the perimeter, then $U_i \ll u_i$ and we are dominated by small polarizations, $l_{ij}=0$. We will not explore this interesting limit here. 
If, on the contrary, we take the perimeter to become null faster than the inner diagonals then $u_i \ll U_i$ and we are dominated by large polarizations. (Then we can simplify the blocks as in the 5pt case, see (\ref{F6check}).) This limit preserves cyclicity and is the most natural one from a WL duality point of view as we are first constructing the polygon boundary and only then considering the simplifying limit $U_i\to 0$.

There is another set of cross-ratios $v_{i}$ ($u_{i,i+1}$ in  \cite{Alday:2010zy})
\begin{equation}
v_i \equiv \frac{x_{i-1,i}^2x_{i+1,i+2}^2}{x_{i-1,i+1}^2x_{i,i+2}^2} \,
\end{equation}
which are completely local and valid for any polygon but contain two vanishing distances (inconvenient when taking a null limit at a time). We will use the \textit{single null distance} cross-ratios~$u_i$ to bootstrap the results which we  present using the \textit{local} cross-ratios~$v_i$.

\section{Bootstrap}
\subsection{Cyclic Correlators}
We consider $n$-pt functions of the lightest single trace operator. Since we are after a full null limit where the various external points approach the vertices of a null polygon, it is convenient to make cyclicity manifest by stripping out a cyclic space-time dependent factor. For example, in planar $\mathcal{N}=4$ SYM we consider n-point functions of $20'$ operators $\mathcal{O}_j\propto \text{tr}( (y_j)_A \phi^A(x_j))^2$ and write 
\begin{equation}
\Big\langle \prod_{i=1}^n \mathcal{O}_j \Big\rangle = \prod_{i=1}^n \frac{y_i \cdot y_{i+1}}{(x_{i}-x_{i+1})^2} \times G_n \label{GnDef}
\end{equation}
Then $G_n$ is a \textit{cyclic} function of the conformal cross-ratios alone \cite{footnote2}. We pick the normalization such that $G_n=1+\mathcal{O}(\lambda)$ at tree level. The simplest cyclic correlator is $G_4$ which in the null square limit was bootstrapped in \cite{Alday:2013cwa}. Cyclicicity ($u \leftrightarrow v$) combined with the large spin limit described in (\ref{ABBeq}) fixes the correlator to be
\begin{align}
G_4 & \simeq {\color{magenta} e^{\sum_{i=1}^4\ln{v_i}\left(\frac{g}{4}-\frac{f}{16}\ln v_{i+1}\right)}} \label{4ptCompact} \\
&\!\!\!\!\!\! \times {\color{blue} (\mathcal{N}^{\circ\circ\bullet})^2 \prod_{i=1}^4 e^{-\frac{4}{f} \frac{\partial^2}{\partial \ln v_i \partial \ln v_{i+1}}} \prod_{i=1}^4\Gamma\Big( 1-\frac{g}{2}+\frac{f}{4}\ln v_i \Big)}\nonumber
\end{align}
where the $v_j$ relate to the conventional 4pt function cross-ratios as $v_1=v_3=v$ and $v_2=v_4=u$ \cite{footnote3}.  

\subsection{5pt function}
The bootstrap of the cyclic 5pt function $G_5$ in the null polygon limit follows closely the 4pt function bootstrap of \cite{Alday:2013cwa}. We plug a perturbative ansatz 
\begin{align}
\hat{P}^{\circ \bullet \bullet} & \simeq \sum_{k=0}^\infty\lambda^k
 \sum_{i_k}  c_{k,i_1,i_2,i_3}\times \ln^{i_1}J_1 \ln^{i_2} J_2 \ln^{i_3} l\;. \label{Peq}
\end{align}
for the relevant combination of structure constants 
\begin{equation}
\hat{P}^{\circ \bullet \bullet}=\hat{C}^{\circ \bullet \bullet}(J_1,J_2,l)\hat{C}^{\circ \circ \bullet}(J_1)\hat{C}^{\circ \circ \bullet}(J_2)\;, \label{PDEF}
\end{equation}
arising in the OPE decomposition of the 5pt function 
\begin{equation}
G_5 \simeq \int_0^\infty \frac{1}{4}\hat{P}^{\circ \bullet \bullet}u_1^{1+\frac{\gamma(J_1)}{2}}u_3^{1+\frac{\gamma(J_2)}{2}}\check{F}_5 \textnormal{d}j_1 \textnormal{d}j_2 \textnormal{d}l\;,
\label{toBootstrap5pt}
\end{equation}
and compute this object to any desired order in perturbation theory \cite{footnoteSumByInt}. The integrals involved are very similar to those arising in the 4pt function case of \cite{Alday:2016mxe}, see appendix \ref{app:Cyclicity}.

In this way we obtain $G_5$ in the polygon limit expressed in terms of the unfixed constants $c_{k,i_1,i_2,i_3}$ in the structure constants (\ref{Peq}) and of the collinear and cusp anomalous dimensions appearing in (\ref{anomalousDimension}). Now, because we picked a particular OPE channel for the decomposition (see figure \ref{5pt limits}b) the whence obtained result will not be automatically cyclic. Imposing cyclicity (i.e. invariance under $u_i \to u_{i+1}$) will thus strongly constrain the unfixed constants $c_{k,i_1,i_2,i_3}$, resulting in
\begin{align}
\hat{C}^{\circ \bullet \bullet} & \simeq  \mathcal{N}^{\circ \bullet \bullet}e^{-\frac{f(\lambda)}{4}(\ln^2 l+\ln 4 \ln (J_1 J_2))-\frac{g(\lambda)}{2}\ln l}\;.
\label{CJJScalar}
\end{align}
This all loop result for the large spin and large polarization behaviour of the OPE coefficients involving two spinning operators is the counterpart of (\ref{ABBeq}) with a single spinning operator \cite{ft4}. Having completely constrained the structure constants in the OPE decomposition, we can then write down the correlator itself as 
%
\begin{align}
G_5 & \simeq  {\color{purple}e^{\frac{-f}{16}\left(\sum_{i=1}^5\ln v_{i}\right)^2+\sum_{i=1}^5 \ln v_{i}\left(\frac{g}{4}+\frac{f}{4}\ln v_{i+2}\right)}}  \times \label{5ptCompact} \\
&\!\!\!\!\!\!\!\!\!\! \times {\color{blue} \mathcal{N}^{\circ\bullet\bullet}(\mathcal{N}^{\circ\circ\bullet})^2 \prod_{i=1}^5 e^{-\frac{4}{f} \frac{\partial^2}{\partial \ln v_i \partial \ln v_{i+1}}} \prod_{i=1}^5\Gamma\Big( 1-\frac{g}{2}+\frac{f}{4}\ln v_i \Big)}\nonumber
\end{align}
We interpret it physically in the next section. We compared our results with perturbative one-loop results in the literature. In the limit of large spin and polarization, the correlation function (\ref{5ptCompact}) agrees with the one-loop computations of \cite{Drukker:2008pi} and the structure constant (\ref{CJJScalar}) agrees with the perturbative computations of \cite{Bianchi:2019jpy}. 
 


\subsection{6pt function}

For the 6pt function we focus on the origin limit, where we plug the ansatz 
\begin{align}
\hat{P}^{\bullet \bullet \bullet} & \simeq \sum_{k=0}^\infty\lambda^k
 \sum_{i_k}  c_{k,i_1,\dots,i_6} \prod_{n=1}^3  \ln^{i_n}J_n\ln^{i_{n+3}} l_{4-n} \, 
\label{Phex}
\end{align}
($i_1+ i_2  + i_3\le k, i_1+\dots+i_6  \le 2k$) for the relevant combination $\hat{P}^{\bullet \bullet \bullet} = \hat{C}^{\bullet \bullet \bullet}(J_1,J_2,J_3,l_1,l_2,l_3) \prod_{j=1}^3  \hat{C}^{\circ \circ \bullet}(J_j)$ appearing in the OPE decomposition
\begin{align}
G_6 \simeq \int_0^\infty \frac{1}{8}\hat{P}^{\bullet \bullet \bullet}\,u_1^{1+\frac{\gamma(J_1)}{2}}u_3^{1+\frac{\gamma(J_2)}{2}}u_5^{1+\frac{\gamma(J_3)}{2}}\check{F}_6 \prod_{i=1}^3\textnormal{d}j_i  \textnormal{d}l_i  \,.
\label{toBootstrap6pt}
\end{align}
where the block is given by (\ref{F6check}) and we impose cyclicity (i.e. invariance under $u_i\to u_{i+1}$ and $U_i \to U_{i+1}$). At loop order $k$, we find that all power of logs in~(\ref{Phex}) from degree $k$ up to $2k$ are fixed while the remaining constants are such that only the three variables $\mathbb{L}_j \equiv 2\log(J_j)-\log(l_{j-1})-\log(l_{j-2})$ show up. That is, cyclicity fixes the large spin and large polarization structure constant to be
\begin{align}
    \hat{C}^{\bullet \bullet \bullet} & \simeq 
	 \mathcal{N}^{\bullet \bullet \bullet}  e^{\sum_{i=1}^3\frac{f(\lambda)}{2}\ln J_i \ln\left(\frac{J_i l_i}{2 l_{i+1}l_{i+2}}\right)}     \label{CJJJFinal} \\ 
    &\times \exp\Big({-\frac{f(\lambda)}{8} \sum_{i=1}^3 (\mathbb{L}_j)^2 +\sum_{k\ge 1} \lambda^{k} P_k(\mathbb{L}_1,\mathbb{L}_2,\mathbb{L}_3)}\Big) \nonumber
\end{align}
which translates into the associated 6pt function in the origin limit as
\begin{align}
\hat{G}_6 & \simeq {\color{purple} e^{\sum_{i=1}^6 \frac{f}{16}\ln v_{i}\ln v_{i+3}-\frac{f}{8}\ln v_{i}\ln v_{i+1} +\frac{g}{4} \ln v_i}} \label{6ptCompact}\\
&\!\!\!\!\!\!\!\!\!\! \times {\color{blue} \mathcal{N}^{\bullet\bullet\bullet}(\mathcal{N}^{\circ\circ\bullet})^3 \prod_{i=1}^6 e^{-\frac{4}{f} \frac{\partial^2}{\partial \ln v_i \partial \ln v_{i+1}}} \prod_{i=1}^6\Gamma\Big( 1-\frac{g}{2}+\frac{f}{4}\ln v_i \Big)}\nonumber \\ 
&\!\!\!\! \!\!\!\! \times {\color{ForestGreen} e^{-\frac{f(\lambda)}{8} \sum_{i=1}^3 (\log U_j)^2 +\sum_{k\ge 1} \lambda^{k} P_k(\log U_1,\log U_2,\log U_3)}} \nonumber
\end{align}
where $P_k$ are undetermined totally symmetric polynomials of three variables of degree $k$. In the original ansatz (\ref{Phex}) we had polynomials of twice as many variables and twice the degree, see table~\ref{tableCounting}. 

\begin{table}[t]
  \begin{center}
    \begin{tabular}{c|c|c} 
\textbf{Loop} &     \textbf{Perturbative} &    \textbf{After} \\
\textbf{ Order } &      \textbf{ Ansatz (\ref{Phex}) } &   \textbf{ Cyclicity } \\
\hline
1&      8 &    2  \\
2&      53 &   4 \\
 3&     243 &   8 \\
 4&      708 &     13 \\
  5&         1862 &      20 
    \end{tabular}
        \caption{Number of free parameters in $C^{\bullet \bullet \bullet}$ at large $J_i, l_{ij}$.}
    \label{tableCounting}
  \end{center}
\end{table}


\section{WL Relation}
In the null polygon limit all correlation functions bootstrapped here can be compactly written as~\cite{Alday:2010zy}
\begin{equation}
G_n={\color{purple}(\texttt{Sudakov})_n} \times {\color{blue}(\texttt{J Recoil})_n} \times {\color{ForestGreen} \mathbb{W}_n} \,.
\end{equation}
The first factor contains the leading divergences of the correlator developed as it approaches the null polygon limit. These physical divergences are identified with UV cusp divergences of the dual null polygonal WL~\cite{Alday:2010zy} or IR divergences of the dual gluon scattering amplitudes~\cite{Eden:2010zz} (in $\mathcal{N}=4$ SYM). This Sudakov factor is given in magenta in first lines in our expressions~(\ref{4ptCompact}), (\ref{5ptCompact}) and (\ref{6ptCompact}) and precisely matches with the prediction of \cite{Alday:2010zy}.

The second factor should be identified with a subtle backreaction effect: as the correlation function insertions become null separated, fast particles will propagate between consecutive insertions thus dynamically generating a WL. However, these charged particles interact with each other and get pulled towards each other producing a recoil effect which needs to be taken into account. This $J$ recoil factor is written in blue in the second line in our expressions (\ref{4ptCompact}), (\ref{5ptCompact}) and (\ref{6ptCompact}). It was given in \cite{Alday:2010zy} at two loops and derived to all loops for $n=4$ by Alday and Bissi  \cite{Alday:2013cwa}. Here we derived its all loop expression for $n=5$ and $n=6$; our results agree with the two loop predictions of \cite{Alday:2010zy}. Note that the $J$ recoil factor and the Sudakov factor only depend on the local cross-ratios $v_{i}$ which go to zero in the null polygon limit. The $n>6$ extrapolation of our findings seems obvious as well.

Finally we have the renormalized conformal WL factor $\mathbb{W}_n$ which is finite as we approach the null limit. It is equal to $1$ for $n=4,5$ since there are no cross-ratios to be constructed out of null squares or pentagons while for $n=6$ it only depends on the three finite cross-ratios $U_i$ and is directly related to the hexagon null WL or six point MHV gluon scattering amplitude in planar $\mathcal{N}=4$ SYM.  (In the notation of \cite{origin}, we expect $\mathbb{W}_6=e^{2\mathcal{E}}$.) This WL factor is given (in green) in the third line in (\ref{6ptCompact}) in the limit where all $U_i$ are taken to be very small. Now, in the recent work of \cite{origin} it was observed that in this limit the WL expectation value in planar $\mathcal{N}=4$ SYM exponentiates into a simple quadratic form in $\ln(U_j)$ with coefficients given in terms of two functions $\Gamma_\text{oct}$ and $\Gamma_\text{hex}$ which are explicitly given \cite{Frank}. We conclude that our symmetric polynomials are simply quadratic and read
\begin{eqnarray}
\sum_l \lambda^l P_l &=& 2C_0+\frac{3\Gamma_\text{cusp}-\Gamma_\text{oct}-2\Gamma_\text{hex}}{12}\sum_{j=1}^3\ln^2(U_j) \nonumber \\&& \,\,\,\, -\frac{\Gamma_\text{oct}-\Gamma_\text{hex}}{12}\sum_{j\neq k}\ln U_k \ln U_j\; , \label{PFinal}
\end{eqnarray}
thus fixing the large spin, large polarization structure constants $C^{\bullet\bullet\bullet}$ to all loop orders in this gauge theory. (Here $\Gamma_\text{cusp}=f/2$ is the \textit{fundamental} cusp anomalous dimension.)

\section{Conclusion}
We explored full light-cone OPE limits in 5 and 6-point functions in large $N$ gauge theories. In these limits, the operators approach the vertices of null polygons and the correlators are dominated by the exchange of single trace leading twist operators with large spin.

Five points correlation functions (and corresponding large spin structure constants $C^{\bullet\bullet\circ}$) can be fully determined to all loops in this limit \cite{footnote}.
Six point functions around the origin limit (and corresponding structure constants $C^{\bullet\bullet\bullet}$ involving large spins and polarizations) are strongly constrained -- see table \ref{tableCounting} -- in generic gauge theories and totally fixed in $\mathcal{N}=4$ super Yang-Mills through the relation to the origin limit of null polygonal WLs \cite{origin}.

We believe we are only scratching the surface of a very promising precise connection between (integrated) correlation functions and (integrated) WL. Would be very interesting, for instance, to consider a systematic expansion of correlation functions around the large spin limit and large polarizations considered here and see how this related to the expansion around the origin in the WL side. And of course, there is a plethora of physical WL/amplitudes limits one could consider and translate for the correlation function side.

We explored the conformal bootstrap for six point functions by OPE expanding the external operators in pairs and then considering the resulting spinning three point function. This is the so-called \textit{snowflake}~decomposition \cite{Fortin:2019zkm}. Another OPE decomposition is the so-called \textit{comb} decomposition where OPEs are taken in a sequential way. Conformal block in this limit can be found in \cite{Rosenhaus:2018zqn,Fortin:2019zkm,Parikh:2019dvm}. Could further constraints from the comb decomposition fix the correlator in the origin limit or perhaps shed light on the quadratic truncation in $\log U_i$ found in \cite{origin}? {The comb decomposition reminds of the POPE decomposition \cite{Basso:2013vsa} where each pentagon is concatenated after another; it might  be the natural decomposition to recover the WL collinear limit.}


We hope these games will also pave the way towards a unified integrability description of open and closed strings in AdS/CFT since they relate closed string scattering (correlation functions) to open strings partition functions (WL). Recovering the all loop large spin three point functions results derived here through the hexagon formalism \cite{hexagons} should be very illuminating in this regard. 
Of course, to make progress we will also need more perturbative data.  In the open string pentagon OPE program \cite{Basso:2013vsa}, for instance, such perturbative data was key in validating several integrability based conjectures, see e.g. \cite{dataOPE}. Hexagonal and heptagonal WLs have been computed to mind-blowing perturbative orders, see \cite{Caron-Huot:2020bkp} for a recent review. The same can not be said about higher point correlation functions. 
Apart from some important $n=5$ two loop integrand results in \cite{Eden:2010zz} and the one loop integrated correlators in \cite{Drukker:2008pi}, virtually nothing is known about $n>4$ correlators. 
The 2 loop integrated 6pt function, for instance, would be extremely useful for checking any integrability/bootstrap based formulation since at 1 loop things are often misleadingly simple.

It would be interesting to study the case where the OPE is dominated in some channels by a finite number of exchanged operators. This would be a generalization of \cite{Fitzpatrick:2012yx,Komargodski:2012ek} to higher a number of points.

The exploration of null pentagons and hexagons provides us with a plethora of exact results which seem hardly attainable through the four point function alone. Mia's advice \textit{don't be a square} seems to pay off.

\begin{acknowledgments}
We would like to specially thank Jacob Abajian for collaboration at the initial stages of this project. We thank F.~Alday, A.~Homrich and G.~Korchemsky for comments on the draft. Research at the Perimeter Institute is supported in part by the Government of Canada through NSERC and by the Province of Ontario through MRI. This work was addition- ally supported by a grant from the Simons Foundation (PV: \#488661).
The work of V.G. was supported by by the Coordenacao de Aperfeicoamento de Pessoal de Nivel Superior - Brasil (CAPES) and the Simons Foundation grant 488637 (Simons Collaboration on the Non-perturbative bootstrap). Centro de Física do Porto is partially funded by Fundação para a Ciência e a Tecnologia (FCT) under the grant UID-04650-FCUP. The work of C.B. was suported by Process $n^o$ 2018/25180-5, Fundação de Amparo à Pesquisa do Estado de São Paulo (FAPESP). 
\end{acknowledgments}

\newpage
\appendix

\section{More on Blocks} \label{BlocksAppendix}
The tensors in (4) read \textcolor{black}{$H_{ij}=\epsilon_{i}\cdot x_{ij}\epsilon_{j}\cdot x_{ij}-\frac{1}{2}\epsilon_i\cdot\epsilon_jx_{ij}^2$} and $V_{i,jk}=(\epsilon_{i}\cdot x_{ik}x_{ij}^2-\epsilon_{i}\cdot x_{ij}x_{ik}^2)/x_{jk}^2$. As mentioned above, we can use the light cone OPE (3) multiple times inside a correlation function to obtain a multipoint block
\begin{align}
\langle \prod_{i=1}^{3+n}\mathcal{O}_i\rangle &= \sum_{k_i}\prod_{i=1}^nC_{k_i}\frac{(x_{2i-1 \,2i}\cdot \partial_{\epsilon})^{J_i}}{(x_{2i-1 \,2i}^2)^{\frac{2\Delta_{\phi}-\tau_i}{2}}}\,\!_1F_1(2i-1)\langle \prod_{i=1}^{3}\tilde{\mathcal{O}}_i\rangle\nonumber\\
&= \prod_{i=1}^{n+3}\frac{1}{(x_{ii+1}^2)^\frac{\Delta_\phi}{2}}\prod_{i=1}^nu_{2i-1}^{\frac{\tau_i}{2}}\,\,F_{3+n}\label{eq:lightcoenOPEE2}
\end{align}
where $\tilde{\mathcal{O}}_j$ is either an external or an exchanged operator weather it arose from an OPE. $C_i$ is the OPE coefficient,  $\,_1F_1(i)$ is short notation to denote the hypergeometric function in (3) associated with the OPE in the points $i$ and $i+1$ and the sum over $k$ is the usual sum over primaries.
We use an integral representation for $\!_1F_1$
\begin{align}
\,_1F_1(a,b,z) = \frac{\Gamma(a+b)}{\Gamma(a)\Gamma(b)}\int_0^1dt t^{a-1} (1-t)^{b-1}e^{z\,t} 
\end{align}
to simplify further and obtain the representation (5) for the conformal blocks. 
The measure is $[dy_i] = \frac{\Gamma(\Delta_i+J_i)}{\Gamma^2(\frac{\Delta_i+J_i}{2})}(y_i(1-y_i))^{\frac{\Delta_i+J_i}{2}-1}dy_i$ and  
the prefactors are
\begin{align}
\texttt{kin}_5 &= \left(\frac{u_2u_4}{u_3u_1}\right)^{\frac{\Delta_\phi}{2}}((1-u_2)u_4u_5)^{l}u_{4}^{\frac{\tau_1-\Delta_{\phi}}{2}}u_{5}^{\frac{\tau_2}{2}}\\
\texttt{kin}_6 &= \left(\frac{u_2u_4u_6}{u_1u_3u_5}\right)^{\frac{\Delta_\phi}{2}}\prod_i^3\frac{(\tilde{U}_{i-1}\tilde{U}_{i+1}(\tilde{U}_i-u_{2i}))^{l_{i-1}}}{\tilde{U}_i^{J_i}}\nonumber
\end{align}
where we are using $l_1=l_{23},l_2=l_{13},l_3=l_{12}$ and $\tilde{U}_1=U_1, \tilde{U}_2=U_3, \tilde{U}_3=U_2$. The integrands read
\begin{align}
&\texttt{int}_{5} = (1-y_2-u_5+y_2(u_4+u_5)-y_2u_2u_4u_5)^{J_1-l} \times \nonumber\\
&\frac{(1-y_1+y_1u_5-u_4+u_4y_1-y_1u_2u_4u_5)^{J_2-l}}{(1-\sum_{i=1}^2y_i(1-u_{6-i})+y_1y_2(1+u_2u_4u_5-u_4-u_5))^{a}}\nonumber\\
&\texttt{int}_{6} = \prod_{i=1}^3 \mathcal{A}_{y_{i},y_{i+1}}(\tilde{U}_{i},\tilde{U}_{i-1},u_{2i})^{o_{i-1}}\nonumber\\&\times\mathcal{B}_{y_i,y_{i-1}}(\tilde{U}_{i+1},\tilde{U}_i,\tilde{U}_{i-1},u_{2i+2},u_{2i})^{n_{i+1}}\
\end{align}
with $2a={\sum_{i}(J_i+\Delta_i)-\Delta_\phi}$,
\begin{align}
&\mathcal{A}_{t,y}(X,Y,u)=1+t(1-y)uY/X-t(1-yY)
\\
&\mathcal{B}_{q,y}(X,Y,Z,u,w)=X(1-q)(1-yY)\\
&-uY(1-q)(1-y)-q(1-y)uwZ+qXZ(1-yw)\nonumber
\end{align}
and 
\begin{align}
&2o_j={2\Delta_j-2l_j-\sum_i(J_i-2l_i+\Delta_i)}§,\,n_j=J_j-\sum_{i\neq j}l_{i}\nonumber.
\end{align}
A detailed derivation is in the \texttt{Mathematica} notebook attached with the submission of this publication. 

This representation for the conformal blocks on the light-cone can be checked by Casimir equation of the conformal group. For the OPE channel $(ij)$ it reads \cite{Dolan:2003hv}
\begin{align}
\bigg[\frac{1}{2}(J_{i,AB}+J_{j,AB})&(J_{i}^{AB}+J_{j}^{AB})  - C_{\Delta,J}\bigg]\texttt{pre}_k \,\,G_{\Delta_i,J_i}=0\nonumber
\end{align}
where $J_{i,AB}$ are the generators for the conformal group at position $i$, $C_{\Delta,J}=\Delta(\Delta-d)+J(J+d-2)$ is the Casimir eigenvalue and $\texttt{pre}$ is a prefactor 
\begin{align}
\texttt{pre}_4 &=\frac{1}{(x_{12}^2x_{34}^2)^{\Delta_\phi}} \,, \,\,\,\,\, \texttt{pre}_5 =\frac{1}{(x_{12}^2x_{34}^2)^{\Delta_\phi}} \left(\frac{x_{13}^2}{x_{15}^2x_{35}^2}\right)^\frac{\Delta_\phi}{2}\nonumber\\
\texttt{pre}_6 &=\frac{1}{(x_{12}^2x_{34}^2x_{56}^2)^{\Delta_\phi}} \nonumber
\end{align}
that takes into account the weight of the external operators. These prefactors are more natural to decompose in a particular OPE channel. We decided not to use them in the main text because cyclicity would not be as nice. The Casimir differential equations are lengthy and so we decided to put them in a \texttt{Mathematica} notebook together with the check of the integral representation. Let us also point out that the integral representation is consistent with the $5$-pt conformal blocks derived in \cite{Goncalves:2019znr}. 

The four point block is of course very well known while the five and six blocks are new as far as we know. 

%
%

We will sketch the main idea of how the formulas (8,9) for the conformal blocks at large can be derived. One way to do it is to start from the integral representation series expand around in some cross ratios. The expression that results from this is given in terms hypergeometric function $\,_2F_1$. The large spin expression can then be obtained by approximating the hypergeometric  for a Bessel function. Then we just have to do the sums coming from the series expansion. We attach a Mathematica notebook with this derivation.  Another way to derive (8,9) is to start with the integral representation, take the limits inside the integral and see what is the region of integration that dominates. Finally, these expressions can be checked by the Casimir differential but this does not fix the normalization which is important for the bootstrap. 

At tree level we have
\begin{align}
&P^{\circ \circ \bullet}_{\textrm{tree}}=\frac{2(J_1!)^2}{(2J_1)!}\,, \qquad P^{\circ \bullet \bullet}_{\textrm{tree}} = \prod_{i=1}^2 \frac{2(J_i!)^3}{l!(2J_i)!(J_i-l)!} \,,\label{CTree5}\\
&P^{\bullet \bullet \bullet}_{\textrm{tree}}  = \prod_{i=1}^3 \frac{2(J_i!)^3}{(l_i!)^2(2J_i)!(J_i+l_i-\sum_jl_j)! }.\label{CTree6}
\end{align}
By {multiplying} these tree-level results by the corresponding blocks we can able to use the normalized structure constants $\hat{P}$ when bootstrapping the five and six  point correlation functions in (15) and (19). These normalized blocks, in the large $J_i$ and $l_i$ limit, simplify into
\begin{equation}
\check{F}_5  =\left(\frac{u_2 u_4 u_5}{u_1 u_3}\right) \frac{J_1J_2}{l^2/4}  \prod_{i=1}^2 \Big(\frac{4 l}{J_{i-1}} \Big)^{\frac{\gamma_i}{2}}   \, e^{-l u_2 - \frac{J_2^2 u_4}{l}- \frac{J_1^2 u_5}{l}}
\label{F5check} \,,
\end{equation}
\begin{align}
\check{F}_6 & = \prod_{i=1}^3 \Big{(}2^{\frac{3}{2}+\gamma_i}J_i^{1+2\max\{l_{i-1},l_{i}\}+\frac{\gamma_{i-1}-\gamma_{i+1}}{2}}l_i^{-1-2l_i} u_{2i}^{1-l_{i-1}}\nonumber\\
&\!\!\! \!\!\!|l_{i-1}{-}l_{i}|^{-\frac{1}{2}-|l_i-l_{i-1}|+\frac{\gamma_{i-1}-\gamma_i}{2}}U_{2-i}^{\min\{l_{i-1},l_{i}\} + \frac{\gamma_{i}-\gamma_{i-1}}{2}}  \label{F6check} \\ 
&\!\!\! \!\!\! e^{2\max\{l_{i-1},l_{i}\}}\Big{)}\times e^{\frac{J_2^2U_1}{|l_3-l_1|}+\frac{J_1^2U_2}{|l_2-l_3|}+\frac{J_3^2U_3}{|l_1-l_2|}-\frac{l_3 U_1}{u_2}-\frac{l_1 U_3}{u_4}-\frac{l_2 U_2}{u_6}} \nonumber 
\end{align}

\section{Bootstrap details}
\label{app:Cyclicity}
In this section we explicit some of the bootstrap computations that lead to the all loop expressions for the structure constants (16,20) and correlation functions~(17,21).
We begin by writing (15) explicitly
\begin{align}
G_5 & \simeq \sum_{k=0}^L\lambda^k
 \sum_{i_k}  c_{k,i_1,i_2,i_3}\times \mathcal{I}_5 \label{bootstrapExplicitlely}\\ 
\mathcal{I}_5 & = \int_0^\infty \left(\ln^{i_1}J_1 \ln^{i_2} J_2 \ln^{i_3}  u_1^{\frac{\gamma(J_1)}{2}}u_3^{\frac{\gamma(J_2)}{2}}\check{F}_5\right) \textnormal{d}J_1\textnormal{d}J_2\textnormal{d}l \nonumber
\end{align}
where $\check{F}_5$ is the block multiplied by the tree-level 3pt function, given in (\ref{F5check}).
To evaluate the integrals we expand in $f(\lambda)$ to bring the $\ln J_i$ dependence down from the exponents,
\begin{align}
\mathcal{I}_5 & = \sum_{n,m}^\infty \frac{2^{2g}}{n!m!} \left(\frac{f}{2}\right)^{n+m}l^{-2+g} e^{-l u_2-\frac{J_2^2u_4}{l}-\frac{J_1^2u_5}{l}}\times \nonumber \\
& u_1^{\frac{g}{2}}u_3^{\frac{g}{2}}J_1^{1-\frac{g}{2}}J_2^{1-\frac{g}{2}}\ln^{i_3} l \ln^{i_1+n} J_1 \ln^{i_2+m}J_2\times
\label{I5Mess1}\\
& \left(\ln(4u_2)+\ln l -\ln J_2\right)^n \left(\ln(4u_3)+\ln l-\ln J_1\right)^m \nonumber
\end{align}
Once we change variables to $\mathbb{J}_1=J_1^2u_5/l$, $\mathbb{J}_2=J_2^2 u_4/l$ and $\mathbb{J}_3=lu_2$ all the remaining integrals are of the form
\begin{equation}
\int_0^\infty \left(\mathbb{J}^{t} e^{-\mathbb{J}}\ln^{r}\mathbb{J}\right)  \textnormal{d}\mathbb{J} =  \frac{\partial^r}{\partial \alpha^r}\left(\Gamma\left(1+t+\alpha \right)\right)_{\alpha_i=0} 
\label{ExpInt} \,.
\end{equation}
After integrating we can re-sum the perturbative expansion in $n$ and $m$ obtaining
\begin{align}
\mathcal{I}_5 & = e^{\frac{-f}{8}\left( \ln\left(\frac{16 u_1^2 u_4}{u_2}\right)\ln (u_2 u_5) + \ln\left(\frac{16 u_3^2u_5}{u_2}\right)\ln (u_2u_4)\right)}\times \nonumber \\
& \left(\frac{u_1^2 u_3^2 u_4 u_5 }{2^8 u_2^2}\right)^{\frac{g}{4}} \left( \frac{2^{-2-i_1-i_2}}{u_2 u_4 u_5}\right)(-\ln(u_2)+\partial_3)^{i_3} \times \nonumber \\
& (-\ln(u_2 u_5)+\partial_1+\partial_3)^{i_1} (-\ln (u_2 u_4)+\partial_2+\partial_3)^{i_2}\times \nonumber\\
& \Gamma\left( 1-\frac{g}{4}+\alpha_1+\frac{f}{4}\left(\ln(4u_1u_4)-\partial_2 \right)\right)\times\nonumber \\
& \Gamma\left(1-\frac{g}{4}+\alpha_2+\frac{f}{4}\ln(4u_5 u_3)\right)\times \nonumber \\ 
& \Gamma \left( 1 + \frac{g}{2}+\alpha_3+\frac{f}{4}\left(\ln \left(\frac{16 u_1u_3}{u_2^2}\right)+\partial_3\right)\right)
\label{I5Compact} \,.
\end{align}
We should expand this expression in perturbation theory treating all $\partial_i$ as c-numbers, bring them to the left, act to the right as $\partial_i^n \mathcal{F} \to \frac{\partial^n\mathcal{F}}{\partial \alpha_i^n}$ and  set
$\alpha_i=0$ at the end.

Using this expression we computed the 5pt correlation function up to five loops and imposed cyclicity (i.e. $u_i \to u_{i+1}$) which fixed all the coefficients $c_{k,i_1,i_2,i_3}$, up to an overall spin independent normalization. 
By inspecting (\ref{I5Compact}) we were thus able to guess an all-loop expression for the structure constant (16) such that correlation function $G_5$ is manifestly cyclic invariant, (17). 

For 6pt we proceed in the same fashion, the only caveat being that we started with the conformal blocks (9) given in terms of Bessel functions. The integrals that appear in this case have the same complexity as (\ref{ExpInt}) which lead us to a expression similar to (\ref{I5Compact}). Again by direct inspection of such formula we were able to guess the all loop structure constants (20) and correlation function (21). 

For the five point function we were able to compute the all loop structure constant in a different way. We follow~\cite{Alday:2016mxe} by making a ansatz for the structure constant as a double integral and impose cyclicity direct at the level of the integrand, see \texttt{mathematica} notebook attached.


\begin{thebibliography}{99}
\bibitem{footnoteCNotation}{The superscript of $\mathcal{C}$ can be filled ($\bullet$) or unfilled ($\circ$), associated with protected and unprotected operators, respectively.}
\bibitem{footnote2}{When we take all $x_i$ and $x_{i+1}$ to be null separated we project into the $y$ polarization structure explicitly stripped out in (\ref{GnDef}) here so $G_n$ really only depends on the space-time cross-ratios.}
\bibitem{footnote3}{
We should identify the odd and even $v_j$'s after taking the derivatives.}
\bibitem{footnoteSumByInt}{The replacement of the sums by integrals is justified since in the null polygon limit both the spins and polarizations are very large. The factor of $1/4=(1/2)^2$ accounts for the original sums being only over even spins.}
\end{thebibliography}
\end{document}